%% file: main.tex
\documentclass[sigconf]{acmart} 

\AtBeginDocument{%
}

\copyrightyear{2025}
\acmYear{2025}
\setcopyright{cc}
\setcctype{by}
\acmConference[CIKM '25] {Proceedings of the 34th ACM International Conference on Information and Knowledge Management}{ November 10--14, 2025}{Seoul, Republic of Korea.}
\acmBooktitle{Proceedings of the 34th ACM International Conference on Information and Knowledge Management (CIKM '25), November 10--14, 2025, Seoul, Republic of Korea}
\acmISBN{979-8-4007-2040-6/2025/11}
\acmDOI{10.1145/3746252.3761286}

\usepackage{enumitem}
\usepackage{xspace}
\usepackage{subcaption}
\usepackage{multirow}
\usepackage{colortbl}
\usepackage{array}
\usepackage{booktabs}
\newcolumntype{C}[1]{>{\centering\arraybackslash}p{#1}}
\usepackage{xcolor}
\usepackage{soul}
\usepackage{dsfont}
\usepackage{makecell}
\usepackage{arydshln}
\usepackage{tabularray}


\begin{document}


\title[A Self-Supervised Mixture-of-Experts Framework for Multi-behavior Recommendation]{A Self-Supervised Mixture-of-Experts Framework for Multi-behavior Recommendation}

\settopmatter{authorsperrow=4}
        \author{Kyungho Kim} 
        \orcid{0009-0008-8304-9585}
	\affiliation{%
    	\institution{KAIST}
            \city{Seoul}
            \country{Republic of Korea}
	}
	\email{kkyungho@kaist.ac.kr}

        \author{Sunwoo Kim}
        \orcid{0009-0006-6002-169X}
	\affiliation{%
		\institution{KAIST}
            \city{Seoul}
            \country{Republic of Korea}
	}
	\email{kswoo97@kaist.ac.kr}

        \author{Geon Lee} 
        \orcid{0000-0001-6339-9758}
	\affiliation{%
		\institution{KAIST}
            \city{Seoul}
            \country{Republic of Korea}
	}
	\email{geonlee0325@kaist.ac.kr}
	
	\author{Kijung Shin}
        \orcid{0000-0002-2872-1526}
	\affiliation{%
		\institution{KAIST}
            \city{Seoul}
            \country{Republic of Korea}
	}
	\email{kijungs@kaist.ac.kr}

\begin{abstract}
\input{000abstract.tex}
\end{abstract}

\input{dfn.tex}
\maketitle

\section{Introduction}
\label{sec:intro}
\input{010intro.tex}

\section{Related Work \& Preliminaries}
\label{sec:prelim}
\input{020prelim.tex}

\section{Data Analysis}
\label{sec:analysis}
\input{030analysis.tex}
\section{Proposed Method: \method}
\label{sec:method}
\input{040method.tex}

\section{Experiments}
\label{sec:experiments}
\input{050exp.tex}


\section{Conclusions}
\label{sec:conclusion}
\input{070conclusion.tex}

\section*{Acknowledgements}
This work was partly supported by the National Research Foundation of Korea (NRF) grant funded by the Korea government (MSIT) (No. RS-2024-00406985, 40\%).
This work was partly supported by Institute of Information \& Communications Technology Planning \& Evaluation (IITP) grant funded by the Korea government (MSIT) (No. RS-2024-00438638, EntireDB2AI: Foundations and Software for Comprehensive Deep Representation Learning and Prediction on Entire Relational Databases, 50\%) (No. RS-2019-II190075, Artificial Intelligence Graduate School Program (KAIST), 10\%).

\section*{GenAI Usage Disclosure}
We utilized LLM tools 
to assist with minor writing edits and code debugging. All generated outputs were thoroughly reviewed and revised by the authors to ensure accuracy and integrity.

\bibliographystyle{ACM-Reference-Format}
\balance
\bibliography{ref}



\end{document}

%% file: 000abstract.tex
In e-commerce, where users face a vast array of possible item choices, recommender systems are vital for helping them discover suitable items they might otherwise overlook.
While many recommender systems primarily rely on a user's purchase history, recent \textit{multi-behavior recommender systems} incorporate various auxiliary user behaviors, such as item clicks and cart additions, to enhance recommendations.
Despite their overall performance gains, their effectiveness varies considerably between \textit{visited items} (i.e., those a user has interacted with through auxiliary behaviors) and \textit{unvisited items} (i.e., those with which the user has had no such interactions).
Specifically, our analysis reveals that (1) existing multi-behavior recommender systems exhibit a significant gap in recommendation quality between the two item types (visited and unvisited items) and (2) achieving strong performance on both types with a single model architecture remains challenging.
To tackle these issues, we propose a novel multi-behavior recommender system, \method.
It employs a mixture-of-experts framework, with experts designed to recommend the two item types, respectively.
Each expert is trained using a self-supervised method specialized for its design goal.
In our comprehensive experiments, we show the effectiveness of \method across both item types, achieving up to {65.46\%} performance gain over the best competitor in terms of Hit Ratio@20. 

%% file: dfn.tex
\newcommand{\smallsection}[1]{\vspace{0.2pt}{\noindent {\bf{\underline{\smash{#1}}}}}}
\newtheorem{obs}{\textbf{Observation}}
\newtheorem{prp}{\textbf{Property}}
\newtheorem{dfn}{\textbf{Definition}}
\newtheorem{trm}{\textbf{Theorem}}
\newcommand{\std}{\scriptsize}
\newcommand\red[1]{\textcolor{red}{#1}}
\newcommand\blue[1]{\textcolor{blue}{#1}}
\newcommand\orange[1]{\textcolor{orange}{#1}}
\newcommand\brown[1]{\textcolor{brown}{#1}}
\newcommand\olive[1]{\textcolor{olive}{#1}}
\newcommand\sunwoo[1]{\textcolor{sunwooblue}{#1}}

\definecolor{mygreen}{rgb}{0,0.7,0}
\definecolor{sunwooblue}{RGB}{66, 133, 244}
\newcommand\green[1]{\textcolor{mygreen}{#1}}

\newcommand{\method}{\textsc{MEMBER}\xspace}

\definecolor{verylightgray}{gray}{0.9}

\newcommand{\appropto}{\mathrel{\vcenter{
  \offinterlineskip\halign{\hfil$##$\cr
    \propto\cr\noalign{\kern1pt}\sim\cr\noalign{\kern-1pt}}}}}

\setlength{\textfloatsep}{0.12cm}
\setlength{\dbltextfloatsep}{0.12cm}
\setlength{\abovecaptionskip}{0.12cm}
\setlength{\skip\footins}{0.12cm}

%% file: 010intro.tex
In the era of e-commerce, where an overwhelming number of options are available, users often struggle to discover suitable items, as they are frequently unaware of better choices amid the abundance.
To address this challenge, recommender systems help users identify such items.
By delivering personalized suggestions, these systems not only enhance user experience but also boost retailer sales.

Traditionally, recommender systems have primarily focused on users' purchase history~\cite{koren2008factorization, he2017neural, he2020lightgcn, wang2019neural, zhang2019deep, cai2022lightgcl, gao2023survey, lin2022improving}.
However, it is important to note that users also engage in other behaviors, such as clicking on items and/or adding items to carts. 
These auxiliary user behaviors often provide crucial insights into users' preferences.

Multi-behavior recommender systems~\cite{singh2008relational, loni2016bayesian, jin2020multi, chen2020efficient, chen2021graph, xia2021graph, yan2023cascading, meng2023hierarchical, cheng2023multi, kim2025multi} incorporate these auxiliary behaviors with purchase history to deliver more relevant recommendations for each user.
Notably, multi-behavior recommender systems are demonstrated to provide better recommendations to users compared to those that only rely on user purchase history~\cite{chen2021graph, yan2023cascading, xu2023multi, gu2022self}.

Despite their success, our analysis reveals that the effectiveness of existing multi-behavior recommender systems varies across \textit{visited items}, which a user clicked or added to the cart before purchase, and \textit{unvisited items}, which are directly purchased without prior interactions.
Specifically, our findings indicate that these systems fail to effectively address this disparity.
\textit{First}, the recommendation quality of these models is significantly lower for the unvisited items compared to the visited items.
While they effectively recommend items users have clicked on and/or added to their carts, they often fail to do so for items without prior users' auxiliary behavior.
\textit{Second}, model rankings for visited items differ significantly from those for unvisited items.
Notably, the best-performing model varies across these item types, highlighting the challenge of achieving strong performance in both cases with a single model.

However, an effective multi-behavior recommender system should provide high-quality recommendations for both item types.
The visited items are particularly crucial from a business perspective, as they have a high likelihood of being purchased.
Meanwhile, the unvisited items are especially valuable from a customer perspective, as recommending them helps users discover previously unknown products, thereby broadening their choices.

To address the above challenges and improve recommendation performance across both item types, we propose a novel multi-behavior recommender system,  
 \textbf{\method} (\textbf{\underline{M}}ixture-of-\textbf{\underline{E}}xperts for \textbf{\underline{M}}ulti-\textbf{\underline{BE}}havior \textbf{\underline{R}}ecommendation).
In a nutshell, \method 
employs a mixture-of-experts framework, where each expert is specialized in a distinct recommendation task: (1) recommending visited items and (2) recommending unvisited items.
To achieve this, each expert is trained using a tailored self-supervised learning approach.
Specifically, a \textit{visited-item expert} is trained using specialized contrastive learning to identify auxiliary behaviors that lead to a purchase. 
Meanwhile, an unvisited-item expert combines generative self-supervised learning and contrastive learning to infer purchased items without relying on preceding auxiliary behaviors.

Our comprehensive experiments using four evaluation metrics, nine baseline methods, and three datasets demonstrate the effectiveness of \method in multi-behavior recommendation.
Specifically, it outperforms all baseline methods in widely-used evaluation settings (all 12 settings) and the majority of settings for both visited-item  (5 out of 6 settings) and unvisited-item (4 out of 6 settings) recommendations.
Our key contributions are summarized as:

\begin{itemize}[leftmargin=*]
    \item We empirically show that (1) the recommendation quality of the existing multi-behavior recommender systems drops significantly for unvisited items compared to visited items, and (2) providing accurate recommendations for both item types with a single model architecture is challenging. 
    \item We propose \method, a novel multi-behavior recommender system that tackles this limitation through a mixture-of-expert framework where each expert is specialized in recommending the unvisited items and the visited items, respectively. 
    \item \method achieves a new state-of-the-art in the standard evaluation setting, demonstrating superior performance in both visited-item and unvisited-item recommendations, with up to {65.46\%} gain in {Hit Ratio@20} over the strongest competitor.
\end{itemize}
For \textbf{reproducibility}, we release our code, datasets, and an online appendix at \url{https://github.com/K-Kyungho/MEMBER}.

%% file: 020prelim.tex
In this section, we review related work, introduce key notations, and formally define the problem of multi-behavior recommendation.


\subsection{Related Work}\label{sec:rw}

\smallsection{Multi-behavior recommender systems.} 
Multi-behavior recommendation leverages various types of user-item interactions to improve recommendation accuracy by extracting relevant signals from auxiliary behaviors~\cite{singh2008relational, zhao2015improving, loni2016bayesian, qiu2018bprh, jin2020multi, you2020graph, xia2021graph, wei2022constrastive, guo2023compressed, yan2023cascading, meng2023parallel, lee2024mule}; for a detailed survey, see~\cite{kim2025multi}.
Notably, in various recommendation tasks, multi-behavior recommender systems are demonstrated to be more effective than single-behavior recommender systems (i.e., methods that only leverage users' previous purchase history)~\cite{jin2020multi, chen2021graph, xia2021graph, yan2023cascading}.
Prior works focus on how to effectively model auxiliary behaviors, exploring various techniques, such as transformer- and attention-based approaches~\cite{xia2020multiplex, yan2023mbhgcn, lee2024mule}, self-supervised learning~\cite{yang2021hyper, gu2022self, wei2022constrastive, wu2022multi, xu2023multi} and encoding architecture 
of auxiliary behaviors~\cite{yan2023cascading, cheng2023multi}.

However, despite their use, auxiliary behaviors have not been subjected to in-depth analysis regarding their impact on recommendation performance. Especially, recommendation performances on visited items and unvisited items, two critical aspects for evaluating multi-behavior recommender systems, have not been systematically distinguished (see Section~\ref{sec:analysis}).
As a result, to the best of our knowledge, no approach has been developed to address the distinct challenges of visited and unvisited item recommendations.

To bridge this gap, our work is the first to formally define visited items and unvisited items recommendations in the multi-behavior setting and introduce a Mixture of Experts (MoE)-based framework that can select expert networks based on behavioral signals. Our method is detailed in Section~\ref{sec:method}.

\begin{table}[t]
  \caption{Frequently-used notations}
  \label{tab:notation}
  \centering
  \scalebox{0.87}{
  \begin{tabular}{c l}
    \toprule
    
    \textbf{Notation}& \textbf{Definition} \\
    
    \midrule
    \midrule
    
    $u \in \mathcal{U}$ & User $u$ in the user set $\mathcal{U}$\\
    \midrule
    
    $i \in \mathcal{I}$ & Item $i$ in the item set $\mathcal{I}$\\
    \midrule
    
    $m \in \mathcal{M}$ &  \makecell[l]{Behavior $m$ within the behavior set \\
                        (e.g., click, collect, cart, and buy)} \\
    \midrule
    $\mathcal{E}^{(m)}$ & User-item interaction set corresponding to $m$ \\
    \midrule
    $\mathcal{I}_u^{(m)}$ & Items that user $u$ has interacted corresponding to $m$ \\
    \midrule
    $\mathcal{C}_u^{(V)}$ & Visited items of user $u$\\
    \midrule
    $\mathcal{C}_u^{(U)}$ & Unvisited items of user $u$\\
    \midrule
    $f$ & LightGCN~\citep{he2020lightgcn} graph convolution function \\
    \midrule
    $\mathbf{E}\in \mathbb{R}^{\vert \mathcal{U}\vert \times d}$ 
    & \makecell[l]{User embedding matrix, where $u-$th row \\ 
    corresponds to user $u$'s embedding $\mathbf{E}_{u}$ } \\
    \midrule
    $\mathbf{H} \in \mathbb{R}^{\vert \mathcal{I}\vert \times d}$ & \makecell[l]{
    Item embedding matrix, where $i-$th row \\ 
    corresponds to item $i$'s embedding $\mathbf{H}_{i}$ 
    }\\

    \bottomrule
  \end{tabular}}
  \vspace{5pt}
\end{table}

\smallsection{Mixture-of-experts (MoE) in recommender systems.} 
MoE architectures have been widely employed in various recommendation tasks~\cite{ma2018modeling, ma2019snr, tang2020progressive, li2023adatt, zhang2024m3oe}, particularly in multi-domain and multi-task settings, where they are used to capture domain- or task-specific knowledge. In these applications, the objectives for each domain or task are typically explicit and well-defined, enabling the model to learn distinct representations (or parameters) for each recommendation target. For instance, in multi-domain recommendation, MoE-based approaches allocate separate experts to different product categories, allowing them to specialize in domain-specific characteristics. However, directly adopting these architectures for multi-behavior recommendation poses unique challenges. Unlike multi-domain or multi-task settings where each task has a well-defined objective, multi-behavior recommendation involves multiple behavioral signals as inputs (e.g., clicks, cart additions) but a single final target (e.g., purchase). This structural difference makes it non-trivial to directly apply existing MoE frameworks. 

To address this challenge, our work proposes a new perspective on the formulation of the target task driven by real-world data analysis, where we differentiate between the recommendation of visited items and unvisited items. This reformulation allows MoE to be effectively used for multi-behavior recommendation by assigning specialized experts to capture distinct behavioral signals and their varying contributions to each target recommendation scenario.

\subsection{Notations and problem definition}\label{sec:prelim}
Let $\mathcal{U} = \{u_{1}, \cdots, u_{\vert \mathcal{U} \vert}\}$ denote a set of users and $\mathcal{I} = \{i_{1}, \cdots, i_{\vert \mathcal{I} \vert}\}$ denote a set of items. 
Users interact with items through various types of behaviors (e.g., click, collect, add-to-cart, and buy).
Let $\mathcal{M}=\{m_1, \cdots, m_{|\mathcal{M}|}\}$ represent the set of all possible behaviors (typically, $\mathcal{M}=\{\text{click}, \text{collect}, \text{cart}, \text{buy}\}$). 
Interactions for each behavior $m \in \mathcal{M}$ are represented as a distinct user-item bipartite graph $\mathcal{G}^{(m)} = (\mathcal{V}, \mathcal{E}^{(m)})$, where $\mathcal{V}=\mathcal{U} \cup \mathcal{I}$ is the set of users and items, which is consistent across all behaviors, and $\mathcal{E}^{(m)}$ represents the user-item interactions specific to behavior $m$.
Specifically, $(u,i)\in \mathcal{E}^{(m)}$ indicates that user $u$ has interacted with item $i$ through behavior $m$.
For each user $u$, we use $\mathcal{I}_u^{(m)}$ to denote the set of items that the user interacted through behavior $m$, i.e., $\mathcal{I}_u^{(m)}=\{i\in \mathcal{I} : (u,i)\in \mathcal{E}^{(m)}\}$.
Note that users can interact with the same item through multiple behaviors.
The buying behavior is referred to as the \textit{target behavior}, as predicting the items that users will buy is the goal of multi-behavior recommendation, as formalized below.
The remaining behaviors are referred to as \textit{auxiliary behaviors}, as they provide additional information to help us provide better recommendations to users.
Based on these concepts,
the problem of multi-behavior recommendation is defined as follows:
\begin{dfn}[Multi-Behavior Recommendation]
    Given a set of user-item interactions $\mathcal{E}^{(m)}$ for each behavior $m \in \mathcal{M}$, the goal of multi-behavior recommendation is to provide each user $u\in \mathcal{U}$ with a ranked list (or top-$k$) of items that the user has not yet bought but is likely to buy in the future.
\end{dfn}

%% file: 030analysis.tex




In this section, we first define the concepts of visited and unvisited items, and then we analyze the limitations of the existing multi-behavior recommender systems, focusing on the recommendation quality gap between visited and unvisited items.

\begin{figure}[t]
    \vspace{-1mm}
    \centering
        \includegraphics[width=0.925\linewidth]{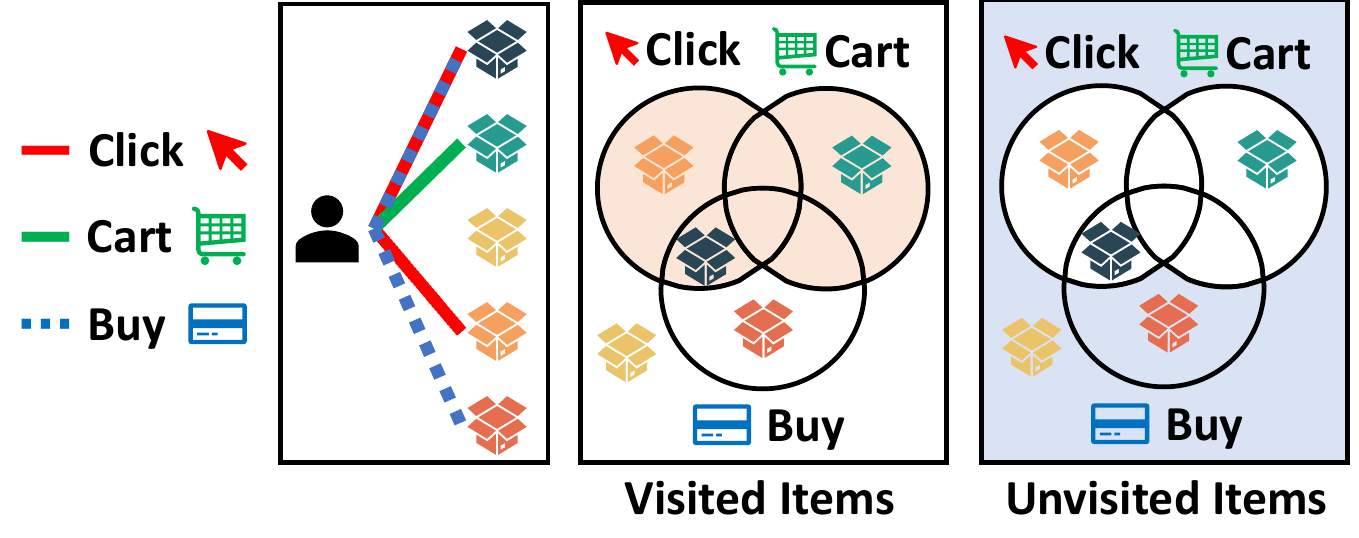}
        \caption{
        An example of the visited items ($\mathcal{C}_u^{(V)}$) and unvisited items ($\mathcal{C}_u^{(U)}$) for user $u$.
        The set of items interacted with through each behavior (e.g., click, cart, buy) is represented as a circle, and these item sets are collectively visualized using a Venn diagram. 
        The shaded areas in each Venn diagram indicate the visited items (left) and unvisited items (right).
        }\label{fig:def}
\end{figure}

\subsection{Analysis Settings}


\smallsection{Visited \& unvisited items (refer to Figure~\ref{fig:def}).}
For each user, the \textit{visited items} are those the user has interacted with through auxiliary behaviors.
Formally, the set of visited items of user $u$, denoted as $\mathcal{C}^{(V)}_{u}$, is defined as 
$\mathcal{C}^{(V)}_{u} = \bigcup_{m\in \mathcal{M}\setminus \{\text{buy}\}} \mathcal{I}^{(m)}_{u}$.
In contrast, the \textit{unvisited items} are those the user has not interacted with through auxiliary behaviors.
The set of the unvisited items of user $u$, denoted as $\mathcal{C}^{(U)}_{u}$, is defined as $\mathcal{C}^{(U)}_{u} = \mathcal{I} \setminus (\bigcup_{m\in \mathcal{M} \setminus \{\text{buy}\}} \mathcal{I}^{(m)}_{u})$.
Note that $\mathcal{C}^{(V)}_{u}$ and $\mathcal{C}^{(U)}_{u}$ are disjoint, i.e., $\mathcal{C}^{(V)}_{u} \cap \mathcal{C}^{(U)}_{u}=\varnothing$. 

\begin{figure}[t]
    \vspace{-1mm}
    \centering
        \includegraphics[width=0.9\linewidth]{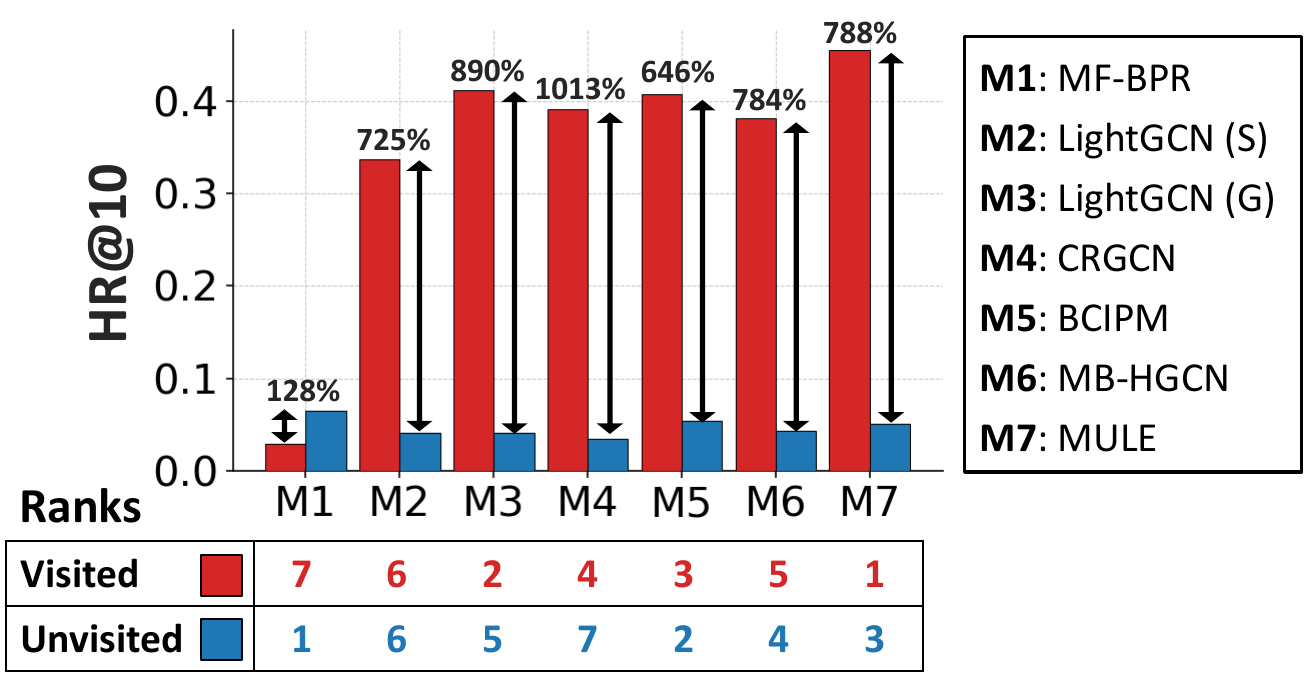}
        \caption{\label{fig:obs}Performance comparison of visited vs. unvisited item recommendation on the Tmall dataset across two single-behavior and five multi-behavior recommender systems. 
        The black arrows denote 
        performance gaps, and ranks (lower is better) indicate model effectiveness in each setting.}
\end{figure}

\smallsection{Difference between cold-start items and unvisited items.} The notion of unvisited items differs from the concept of cold-start items~\cite{gope2017survey, kim2024content, zhou2023contrastive}.
While cold-start items refer to items with few or no interactions at the global level (i.e., across all users), unvisited items are defined at the level of individual users.
Specifically, an unvisited item for a user is one that the user has not interacted with, even though the same item may be a visited item for other users with interaction history.


\smallsection{Model training and evaluation.}
We train the model by using all given behavior information (i.e., $\mathcal{G}^{(m)}, \forall m \in \mathcal{M}$).
Then, we evaluate the recommendation quality of the trained model for (1) visited items and (2) unvisited items, for each user $u$ whose target behavior (i.e., purchase) is in the test set.
To this end, we first categorize the corresponding user's \textit{test} target behavior as either (1) a \textit{visited-item purchase} if the corresponding item belongs to $\mathcal{C}^{(V)}_{u}$, or (2) a \textit{unvisited-item purchase} if the corresponding item belongs to $\mathcal{C}^{(U)}_{u}$.
The trained model's visited-item recommendation quality is evaluated based on its ability to assign high scores to visited-item purchases compared to other items in $\mathcal{C}^{(V)}_{u}$. 
Similarly, its unvisited-item recommendation quality is assessed by how well it assigns high scores to unvisited-item purchases compared to other items in $\mathcal{C}^{(U)}_{u}$.
To measure this, we employ the Hit Ratio of the top 10 scored items (HR@10) for visited-item and unvisited-item recommendation, respectively.
We use three real-world multi-behavior recommendation datasets, Tmall, Taobao, and Jdata, described in Section~\ref{sec:exp:setting}. 
We report results on Tmall in the main paper, and more results are provided in Appendix C~\cite{online2025appendix}.

\subsection{Analysis Results}
We present two key observations derived from our analysis.

\smallsection{(O1) Performance gap between visited and unvisited items.} \\
As shown in Figure~\ref{fig:obs}, all methods undergo a  significant absolute performance drop in unvisited-item recommendation compared to visited-item recommendation.
Specifically, the performance of visited-item recommendation is, on average, 824.2\% higher than that of unvisited-item recommendation across all multi-behavior recommender baselines.
This result indicates that these methods perform significantly worse in recommending unvisited-items compared to visited-items, emphasizing the need for improvements in unvisited-item recommendation. Surprisingly, the simple single-behavior model MF-BPR~\cite{rendle2012bpr}, which only uses the target behavior, achieves the highest performance for unvisited items.

\smallsection{(O2) No one-size-fits-all model.}
As shown in Figure~\ref{fig:obs}, the performance rankings of existing multi-behavior recommender systems for visited-item recommendations differ greatly from those for unvisited-item recommendations.
Especially, the best-performing model differs across the two settings. This discrepancy demonstrates the challenge of achieving strong recommendation performance for both item types using a single recommender system.

%% file: 040method.tex
\begin{figure*}[t]
    \centering
        \includegraphics[width=1.0\linewidth]{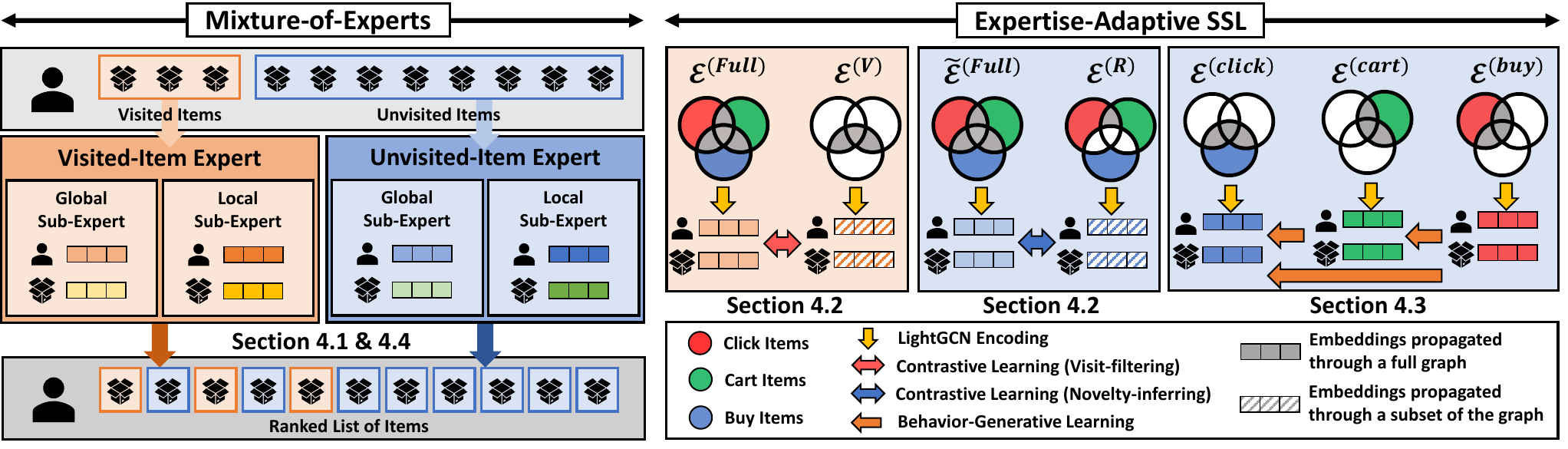}
        \caption{Visualization of \method, the proposed multi-behavior recommendation method based on a mixture of experts.
        \method employs two experts: (1) a visited-item expert, specialized for visited-item recommendations, and (2) an unvisited-item expert, specialized for unvisited-item recommendations.
        This specialization is guided by expertise-adaptive self-supervised learning (SSL), where the visited-item expert leverages visit-filtering contrastive learning, while the unvisited-item expert integrates novelty-inferring contrastive learning and behavior-generative learning.
        } \label{fig:model}
\end{figure*} 


We present \method (\textbf{\underline{M}}ixture-of-\textbf{\underline{E}}xperts for \textbf{\underline{M}}ulti-\textbf{\underline{BE}}havior \\\textbf{\underline{R}}ecommendation), a novel multi-behavior recommender system designed to address the challenges analyzed in Section~\ref{sec:analysis}.
In a nutshell, \method leverages the \textit{Mixture-of-Experts} (MoE) framework, where each expert is respectively specialized for (1) visited-item recommendations and (2) unvisited-item recommendations.
Specifically, to address the challenge of achieving high performance with a single model (O2), each expert is independently parameterized to capture distinct patterns in the data based on its designated objective.
In particular, to mitigate the poor performance in unvisited-item recommendations (O1), we employ a specialized self-supervised learning scheme for the corresponding expert, which is distinct from a separate scheme applied to the visited-item expert. 

Below, we first describe the overall architecture of \method (Section~\ref{sec:method:moe}).
Next, we introduce the specialized self-supervised learning used to train respective experts of \method (Sections~\ref{sec:method:sslcon} and \ref{sec:method:sslgen}). We then detail the final scoring mechanism and training objective (Section~\ref{sec:method:scoring}), followed by complexity analysis (Section~\ref{sec:method:complexity}).

\subsection{Architecture of Each Expert}\label{sec:method:moe}
In this subsection, we describe the sub-module architecture employed by both experts in \method. 
We first formally define the graph convolution function used in our approach and then describe each component of \method step by step based on this function.

\smallsection{Graph convolution function.}
We encode user and item initial embeddings and particular graph topology by using LightGCN~\cite{he2020lightgcn}.
Formally, this process is denoted by $f$, which receives initial user and item embeddings (denoted by $\mathbf{E}^{(0)} \in \mathbb{R}^{\vert \mathcal{U}\vert \times d}$ and $\mathbf{H}^{(0)} \in \mathbb{R}^{\vert \mathcal{I}\vert \times d}$) and a set of edges $\mathcal{E}$ to return encoded user and item embeddings (denoted by $\mathbf{E} \in \mathbb{R}^{\vert \mathcal{U}\vert \times d}$ and $\mathbf{H}\in \mathbb{R}^{\vert \mathcal{I}\vert \times d}$).
Therefore, the overall encoding process is denoted by $f(\mathbf{E}^{(0)},\mathbf{H}^{(0)},\mathcal{E})=(\mathbf{E},\mathbf{H})$.

Within each LightGCN layer, we aggregate the embeddings of neighboring nodes.
Specifically, at the $\ell-$th layer, the user $u$ and item $i$ embeddings, denoted by  $\mathbf{E}_{u}^{(\ell)}$ and $\mathbf{H}_{i}^{(\ell)}$ respectively, are computed as follows:
\begin{equation*}
    \mathbf{E}_u^{(\ell)} = \sum_{i\in \mathcal{N}_u} \frac{1}{\sqrt{|\mathcal{N}_u| |\mathcal{N}_i|}} \mathbf{H}_i^{(\ell-1)}, \quad 
    \mathbf{H}_i^{(\ell)} = \sum_{u\in \mathcal{N}_i} \frac{1}{\sqrt{|\mathcal{N}_i| |\mathcal{N}_u|}} \mathbf{E}_u^{(\ell-1)},
\end{equation*}
where $\mathcal{N}_{u} \subseteq\mathcal{I}$ is the set of items that user $u$ has interacted with, while $\mathcal{N}_{i} \subseteq\mathcal{U}$ is the set of users that have interacted with item $i$.
After message passing, we aggregate the embeddings from each layer to form the final embeddings: 
\begin{equation*}
    \mathbf{E}_u = \frac{1}{L+1} \sum_{\ell=0}^{L} \mathbf{E}_u^{(\ell)},  \quad \mathbf{H}_i = \frac{1}{L+1} \sum_{\ell=0}^{L} \mathbf{H}_i^{(\ell)},
\end{equation*}
where $L$ is the total number of layers.
Here, $\mathbf{E}_{u}$ and $\mathbf{H}_{i}$ corresponds to the $u-$th and $i-$th row of $\mathbf{E}$ and $\mathbf{H}$, respectively.

\smallsection{Input modeling.}
We now describe how we compose the initial embeddings ($\mathbf{E}^{(0)}$ and $\mathbf{H}^{(0)}$) and topology ($\mathcal{G}=(\mathcal{V},\mathcal{E})$).
At this stage, our objectives are twofold: (1) capturing the synergistic interactions across diverse behaviors and (2) preserving the unique information encapsulated within each behavior. 
To accomplish this, we use two sub-experts, which are elaborated on as follows:
\begin{itemize}[leftmargin=*]
    \item \textbf{Global sub-expert:} 
    In this sub-expert, we aim to capture the synergy among all types of user interactions. 
    The \textit{global graph} $\mathcal{G}^{(global)}$ combines all interactions from multiple behaviors into a single graph. 
    That is, $\mathcal{G}^{(global)}=(\mathcal{V}, \mathcal{E}^{(global)}=\bigcup_{m\in \mathcal{M}}\mathcal{E}^{(m)})$.
    In addition, user and item embeddings are parameterized as $\mathbf{E}^{(0),(global)} \in \mathbb{R}^{\vert \mathcal{U}\vert \times d}$ and $\mathbf{H}^{(0),(global)} \in \mathbb{R}^{\vert \mathcal{I}\vert \times d}$, and these embeddings are leveraged coupled with $\mathcal{G}^{(global)}$.
    \item \textbf{Local sub-expert:} 
    In this sub-expert, we aim to capture the unique interaction patterns specific to its corresponding behavior.
    To do so, we use 
    a collection of \textit{local graphs}, each consisting of the interactions associated with a distinct behavior.
    That is, each behavior $m$ is represented by the corresponding local graph $\mathcal{G}^{(m)}$. 
    In addition, user and item embeddings are parameterized as $\mathbf{E}^{(0),(local)} \in \mathbb{R}^{\vert \mathcal{U}\vert \times d}$ and $\mathbf{H}^{(0),(local)} \in \mathbb{R}^{\vert \mathcal{I}\vert \times d}$, and these embeddings are shared across different local graphs. 
\end{itemize}
Note that the parameters are not shared across different experts, to ensure each expert learns distinct information based on their objective.

\smallsection{Encoding.}
Each expert first performs encoding for global and local sub-experts independently.
First, global sub-experts obtain global user and item embeddings simply as follows:
\begin{equation}\label{eq:globalview}
    \mathbf{E}^{(global)}, \mathbf{H}^{(global)} = f(\mathbf{E}^{(0), (global)}, \mathbf{H}^{(0), (global)}, \mathcal{E}^{(global)}),
\end{equation}
Second, local sub-experts compute local user and item embeddings by averaging the embeddings obtained from each local graph (i.e., behavior graph).
Formally, local user and item embeddings, denoted as $\mathbf{E}^{(local)}$ and $\mathbf{H}^{(local)}$, are derived as follows:
\begin{equation}\label{eq:localview}
   \mathbf{E}^{(local)} = \sum_{m \in \mathcal{M}} \frac{\mathbf{E}^{(m),(local)}}{\vert \mathcal{M}\vert}, \mathbf{H}^{(local)} = \sum_{m \in \mathcal{M}} \frac{\mathbf{H}^{(m),(local)}}{\vert \mathcal{M}\vert},
\end{equation}
where 
\begin{equation}\label{eq:localbehavior}
\mathbf{E}^{(m), (local)}, \mathbf{H}^{(m), (local)} = f(\mathbf{E}^{(0), (local)}, \mathbf{H}^{(0), (local)}, \mathcal{E}^{(m)})    
\end{equation}
and $f$ is the aforementioned graph convolution function.

\smallsection{Scoring.} Each expert then computes the recommendation score for user $u$ and item $i$, denoted by $s_{ui}$, by performing a weighted average of the scores produced by the respective sub-experts (i.e., global sub-expert and local sub-expert).
Formally, the score is computed as follows:
\begin{equation}\label{eq:rec_score}
    s_{ui} = \lambda (\mathbf{E}_u^{(global)})^{T} \mathbf{H}_i^{(global)}  + (1 - \lambda) (\mathbf{E}_u^{(local)})^{T} \mathbf{H}_i^{(local)} \in \mathbb{R},
\end{equation}
where $\lambda \in (0, 1)$ is a hyperparameter that adjusts the weighting between scores from the global and local sub-experts.

\subsection{Expertise-Adaptive SSL: (1) Contrastive Learning}\label{sec:method:sslcon}
In this subsection, we propose expertise-adaptive contrastive learning schemes to train two experts for distinct objectives.

\smallsection{Design goals and notations.} The visited-item expert aims to rank items with which a user has already interacted, while the unvisited-item expert aims to discover and rank items that a user has not yet interacted with. To support these distinct objectives, we first partition user-item interactions into two subsets, each relevant to the goal of one expert. Then,
each expert is trained using a contrastive objective, comparing its assigned subset against the full interaction set. This encourages the experts to focus on the portion of interactions most relevant to their task. 

To elaborate further, we first partition the interaction set into $\mathcal{E}^{(V)}$ and $\mathcal{E}^{(R)}$ as follows:
\begin{itemize}[leftmargin=*]
\item $\mathcal{E}^{(V)}$: A \textbf{\underline{v}isited-purchase set} containing only purchases preceded by at least one auxiliary behavior. That is, ${\mathcal{E}}^{(V)} = \mathcal{E}^{(buy)} \cap ( \bigcup_{m \in \mathcal{M} \setminus \{\text{buy}\}}\mathcal{E}^{(m)} ).$
\item $\mathcal{E}^{(R)}$: A \textbf{\underline{r}emaining set}, defined as the complement of $\mathcal{E}^{(V)}$ within the global interaction set, i.e., ${\mathcal{E}}^{(R)} = \mathcal{E}^{(global)} \setminus \mathcal{E}^{(V)}$.  
\end{itemize}
Then, we apply \textit{visit-filtering contrastive learning} for the visited-item expert by aligning (1) embeddings obtained from all interactions and (2) embeddings obtained based on $\mathcal{E}^{(V)}$.
This helps the visited-item expert attend to auxiliary interactions that lead to purchases, which are most directly relevant to its task.
After that, for the unvisited-item expert, we employ \textit{novelty-inferring contrastive learning}, which aligns (1) embeddings obtained from all interactions with (2) embeddings obtained from $\mathcal{E}^{(R)}$. This prevents the unvisited-item expert from being dominated by visited-purchase signals
and helps it attend to 
purchases occurring without any auxiliary actions, which are most relevant to its task.


\smallsection{Visit-filtering contrastive learning.} 
We now describe the details of visit-filtering contrastive learning for the visited-item expert, where all embeddings are derived from this expert.
For the first view, we use the user and item embeddings obtained from Eq.~\eqref{eq:localview}, denoted as $\mathbf{E}^{(full)}$ and $\mathbf{H}^{(full)}$.
For the second view, we derive the user and item embeddings, denoted as $\mathbf{E}^{(V)}$ and $\mathbf{H}^{(V)}$, as follows:
$$\mathbf{E}^{(V)}, \mathbf{H}^{(V)} = f(\mathbf{E}^{(0), (local)}, \mathbf{H}^{(0), (local)}, {\mathcal{E}}^{(V)}),$$
where $f$ is the LightGCN-based graph convolution~\citep{he2020lightgcn}, which is described in Section~\ref{sec:method:moe}.
Finally, we contrast embeddings from these two views (${\mathbf{E}}^{(full)}$, ${\mathbf{H}}^{(full)}$) and (${\mathbf{E}}^{(V)}$, ${\mathbf{H}}^{(V)}$). 
To achieve this, we compute the user and item contrastive losses, denoted by $\mathcal{L}_{\text{CL}}^{\text(V), \text{(user)}}$ and $\mathcal{L}_{\text{CL}}^{\text(V), \text{(item)}}$, respectively, as follows:
\begin{align*}
    \mathcal{L}_{\text{CL}}^{\text(V), ({user})} &= - \sum_{u \in \mathcal{U}} \log \frac{\exp \left( \text{cos} \left( \mathbf{E}_u^{(full)}, \mathbf{E}_u^{(V)} \right) / \tau \right)}{\sum_{u' \in \mathcal{U}} \exp \left( \text{cos} \left( \mathbf{E}_u^{(full)}, {\mathbf{E}}_{u'}^{(V)} \right) / \tau \right)}, \\
    \mathcal{L}_{\text{CL}}^{\text(V), ({item})} &= - \sum_{i \in \mathcal{I}} \log \frac{\exp \left( \text{cos} \left( \mathbf{H}_i^{(full)}, \mathbf{H}_i^{(V)} \right) / \tau \right)}{\sum_{i' \in \mathcal{I}} \exp \left( \text{cos} \left( \mathbf{H}_i^{(full)}, \mathbf{H}_{i'}^{(V)} \right) / \tau \right)},
\end{align*}
where $\tau$ is a temperature hyperparameter and $\text{cos}(\cdot,\cdot)$ represents the cosine similarity. 
The final contrastive loss for the visited-item expert, denoted by $\mathcal{L}_{\text{CL}}^{(V)}$, is computed as:
\begin{equation}\label{eq:clvisited}
    \mathcal{L}_{\text{CL}}^{{(V)}} = \frac{1}{2}\left( \mathcal{L}_{\text{CL}}^{\text(V), ({user})} + \mathcal{L}_{\text{CL}}^{\text(V), ({item})}\right).
\end{equation}

\smallsection{Novelty-inferring contrastive learning.}
We now detail novelty-inferring contrastive learning for the unvisited-item expert, which builds on the intuition introduced above, using embeddings derived from the unvisited-item expert.
For the first view, we use the user and item embeddings from Eq.~\eqref{eq:globalview}, denoted as ${\widetilde{\mathbf{E}}^{(full)}}$ and ${\widetilde{\mathbf{H}}^{(full)}}$. 
For the second view, we compute the user and item embeddings ${\mathbf{E}}^{(R)}$ and ${\mathbf{H}}^{(R)}$ via $f$ as follows:
$${\mathbf{E}}^{(R)}, {\mathbf{H}}^{(R)} = f(\mathbf{E}^{(0), (global)}, \mathbf{H}^{(0), (global)}, {\mathcal{E}}^{(R)}).$$ Finally, we contrast $({\widetilde{\mathbf{E}}^{(full)}}, {\widetilde{\mathbf{H}}^{(full)}})$ with $({{\mathbf{E}}^{(R)}},{{\mathbf{H}}^{(R)}})$ using contrastive losses for users $\mathcal{L}_{\text{CL}}^{{(U)}, \text{(user)}}$ and items $\mathcal{L}_{\text{CL}}^{{(U)}, \text{(item)}}$, which are defined as follows:
\begin{equation*}
    \mathcal{L}_{\text{CL}}^{{(U)}, ({user})} = - \sum_{u \in \mathcal{U}} \log \frac{\exp \left( \text{cos} \left( \widetilde{\mathbf{E}}_u^{(full)}, {\mathbf{E}}_u^{(R)} \right) / \tau' \right)}{\sum_{u' \in \mathcal{U}} \exp \left( \text{cos} \left( \widetilde{\mathbf{E}}_u^{(full)},{\mathbf{E}}_{u'}^{(R)} \right) / \tau' \right)},
\end{equation*}
\begin{equation*}
    \mathcal{L}_{\text{CL}}^{{(U)}, ({item})} = - \sum_{i \in \mathcal{I}} \log \frac{\exp \left( \text{cos} \left( \widetilde{\mathbf{H}}_i^{(full)}, {\mathbf{H}}_i^{(R))} \right) / \tau' \right)}{\sum_{u' \in \mathcal{U}} \exp \left( \text{cos} \left( \widetilde{\mathbf{H}}_i^{(full)}, {\mathbf{H}}_{i'}^{(R)} \right) / \tau' \right)},
\end{equation*}
\noindent
where $\tau'$ is a temperature hyperparameter.
The overall contrastive loss for the unvisited-item expert is defined as:

\begin{equation}\label{eq:clunvisited}
    \mathcal{L}_{\text{CL}}^{{(U)}} = \frac{1}{2}\left( \mathcal{L}_{\text{CL}}^{{(U)}, ({user})} + \mathcal{L}_{\text{CL}}^{{(U)}, ({item})}\right).
\end{equation}

\subsection{Expertise-Adaptive SSL: (2) Generative Learning}\label{sec:method:sslgen}
Although novelty-inferring contrastive learning encourages the unvisited-item expert to capture purchases that occur without auxiliary behaviors, such cases are relatively rare, making it difficult to rely on them alone when training the unvisited-item expert. 
To address this, we supplement novelty-inferring contrastive learning with \textit{behavior-generative learning}, a generative self-supervised approach in which the expert learns to reconstruct dense (i.e., frequent) behaviors from sparse (i.e., rare) ones.\footnote{For consistency across datasets, we use the later-stage signals (e.g., carts, buys) as supervision signals to predict earlier-stage behaviors (e.g., clicks, collects).}
This enables the unvisited-item expert to effectively learn rich representations, especially from sparse behaviors, as required by its task.

Specifically, to formulate the
self-supervised tasks,
we adopt the natural order of behaviors: (1) {clicking on items}, (2) {collecting items}, (3) {adding items to carts}, and (4) {buying items}.
Based on this, we assign an order to each behavior, where an order of behavior $m$ is denoted as $\xi(m)$.
For every ordered pair of behaviors $(m,n)$ with $\xi(m) < \xi(n)$, we train the model to predict edges in $\mathcal{G}^{(m)}$ using propagated embeddings from behavior $n$. To do so, we first obtain the user embeddings $\mathbf{E}_u^{(n), (local)}$ and item embeddings $\mathbf{H}_i^{(n), (local)}$ from Eq.~\eqref{eq:localbehavior}.  
Then, we define the generative loss for the behavior pair $(m,n)$ as follows:
\begin{align*}\label{eq:bceloss}
    \small
    \mathcal{L}_{n\rightarrow m} = - \frac{1}{\vert \mathcal{U}\vert}\sum_{u\in \mathcal{U}} (  \sum_{i \in \mathcal{I}_u^{(m)}} \log \sigma(s_{ui}^{(n)}) +  \sum_{j \notin \mathcal{I}_u^{(m)}} \log \left( 1 - \sigma(s_{uj}^{(n)}) \right) ),
\end{align*}
where $\mathcal{I}_u^{(m)}$ is the set of items connected to user $u$ at behavior $m$, $\sigma$ is the sigmoid function, and $s_{ui}^{(n)}=(\mathbf{E}_u^{(n), (local)}) \left(\mathbf{H}_i^{(n), (local)}\right)^T$.~\footnote{In practice, instead of using all $j \notin \mathcal{I}^{(m)}$ as negative samples, we sample a certain number of items as negative samples for each training epoch.} 
Lastly, we average $\mathcal{L}_{n\rightarrow m}$ over every ordered pair $(m,n)$ with $\xi(m) < \xi(n)$ to define the final generative loss $\mathcal{L}^{\text(U)}_{\text{GEN}}$ as:
\begin{equation}\label{eq:genunvisited}
    \mathcal{L}^{\text(U)}_{\text{GEN}} = \frac{2}{|\mathcal{M}|(|\mathcal{M}|-1)}\sum_{m\in \mathcal{M}} \sum_{n\in \mathcal{M}, \xi{(m)} < \xi{(n)}} \mathcal{L}_{n\rightarrow m}.
\end{equation}

\subsection{All Together: \method}\label{sec:method:scoring}
Finally, we first describe how \method combines the scores of the visited-item expert and the unvisited-item expert to compute the final recommendation score for a user-item pair, and then we describe the final training objective of \method.

\smallsection{Combination of experts with hard gating.}
To integrate experts, typical Mixture-of-Experts (MoE) models typically use \textit{soft gating}, where a small gating network assigns probability weights to each expert, and the final score is a weighted sum of all expert outputs.
However, this approach increases inference costs because every expert must be evaluated for every user-item pair.
To address this, we adopt a \textit{hard gating} mechanism, selecting exactly one expert per user-item pair based on whether the user has previously interacted with the item.
Surprisingly, even though we use only a single expert at a time, our method achieves higher accuracy than alternative strategies, including averaging or learnable soft gating functions, as detailed in Appendix D.2 \cite{online2025appendix}.
This is because visited items and unvisited items are clearly distinguished in our setting, leaving no ambiguity in expert selection.

In order to describe the combination of experts,
we use $s_{ui}^{(V)}$ and $s_{ui}^{(U)}$ to denote the scores from the visited-item expert and unvisited-item expert, respectively (Eq.~\eqref{eq:rec_score}). 
Then, the final score $s_{ui}^*$ for item $i$ to user $u$ is defined as follows:
\begin{equation}\label{eq:finalscore}
s^{*}_{ui} = \mathds{1}[i \in \mathcal{C}_u^{(V)}]s^{(V)}_{ui} + \mathds{1}[i \in \mathcal{C}_u^{(U)}] s^{(U)}_{ui},
\end{equation}
where $\mathds{1}[\texttt{cond}]$ is an indicator function that returns 1 if \texttt{cond} is true and 0 otherwise. 
That is, if item $i$ is a visited item for user $u$, we use $s^{(V)}_{ui}$, and otherwise, we use $s^{(U)}_{ui}$. 


\smallsection{Training of \method.} 
For the final training objective of \method, we incorporate the Bayesian Personalized Ranking (BPR)~\cite{rendle2012bpr} loss, $\mathcal{L}_{\text{BPR}}$, with expert-specific self-supervised losses, $\mathcal{L}_{\text{SSL}}^{{(V)}}$ and $\mathcal{L}_{\text{SSL}}^{{(U)}}$. These self-supervised losses are defined as:
\begin{equation}\label{eq:sslvisited}
    \mathcal{L}_{\text{SSL}}^{{(V)}} = \gamma_1 \cdot \mathcal{L}_{\text{CL}}^{{(V)}},
\end{equation}
\begin{equation}\label{eq:sslunvisited}
    \mathcal{L}_{\text{SSL}}^{{(U)}} = \gamma_2 \cdot \mathcal{L}_{\text{CL}}^{{(U)}} + \gamma_3 \cdot \mathcal{L}_{\text{GEN}}^{{(U)}},
\end{equation}
where $\gamma_1, \gamma_2, \gamma_3 \in \mathbb{R}_{\geq 0}$ are hyperparameters controlling the contribution of each loss component.
The BPR loss  is defined as:
\begin{equation}\label{eq:bpr}
     \mathcal{L}_{\text{BPR}} = - \sum_{u \in \mathcal{U}} \sum_{i\in \mathcal{I}_u^{(\text{buy})}} \sum_{j\notin \mathcal{I}_u^{(\text{buy})}} \log \left( \sigma\left( s^{*}_{ui} - s^{*}_{uj} \right) \right),
\end{equation}
where $\sigma(\cdot)$ denotes the sigmoid function, and $s^{*}_{ui}$ is the recommendation score for user $u$ and item $i$ (Eq.~\eqref{eq:finalscore}).

Building on the above, the final training objective for the visited-item expert is:
\begin{equation}\label{eq:expert_v}
    \mathcal{L}^{({V})} = \mathcal{L}_{\mathrm{BPR}}
    +
    \mathcal{L}_{\mathrm{SSL}}^{({V})}
\end{equation}
Note that only the parameters 
of the visited-item expert are updated via gradient descent to minimize $\mathcal{L}^{({V})}$. 

Similarly, the final training objective for the unvisited-item expert is formulated as 
\begin{equation}\label{eq:expert_u}
    \mathcal{L}^{({U})} =   \mathcal{L}_{\mathrm{BPR}} + \mathcal{L}_{\mathrm{SSL}}^{({U})},
\end{equation}
and only the parameters of the unvisited-item expert are updated via gradient descent to minimize $\mathcal{L}^{({U})}$.

In this way, both experts share a common ranking supervision through $\mathcal{L}_{\mathrm{BPR}}$, while also receiving a specialized self-supervision signal via their respective SSL losses ($\mathcal{L}_{\mathrm{SSL}}^{({V})}$ and $\mathcal{L}_{\mathrm{SSL}}^{({U})}$). 

\subsection{Complexity Analysis}\label{sec:method:complexity}

In this subsection, we analyze the complexity of \method, focusing on its encoding step.
Recall that \method receives as input the global graph $\mathcal{G}^{(global)}=(\mathcal{V}, \mathcal{E}^{(global)}=\bigcup_{m\in \mathcal{M}}\mathcal{E}^{(m)})$ and the local graph $\mathcal{G}^{(m)}$ for each behavior $m$.
Then, it encodes each graph separately with LightGCN~\cite{he2020lightgcn}.
Since the encoding time complexity of LightGCN is $O(\vert \mathcal{V}\vert  + \vert \mathcal{E} \vert)$ for a graph $\mathcal{G} = (\mathcal{V}, \mathcal{E})$~\cite{lee2024revisiting}, encoding all graphs by \method has the following complexity:
\begin{align}
    &O\left(\vert \bigcup_{m \in \mathcal{M}} \mathcal{V}^{(m)}\vert + \vert \bigcup_{m \in \mathcal{M}} \mathcal{E}^{(m)}\vert\right) + \sum_{m \in \mathcal{M}} O(\vert \mathcal{V}^{(m)}\vert + \vert \mathcal{E}^{(m)}\vert), \nonumber \\
    &=O\left(\sum_{m \in \mathcal{M}} \vert \mathcal{V}^{(m)} \vert + \vert \mathcal{E}^{(m)}\vert \right) \label{eq:complexitybound1}.
\end{align}
After encoding all graphs, \method averages the embeddings from the local graphs, whose complexity is
\begin{equation}
O\left(\vert \bigcup_{m \in \mathcal{M}} \mathcal{V}^{(m)}\vert + \vert \bigcup_{m \in \mathcal{M}} \mathcal{E}^{(m)}\vert\right).  \label{eq:complexityaverage}  
\end{equation}
Since the complexity in Eq.~\eqref{eq:complexitybound1} dominates that in Eq.~\eqref{eq:complexityaverage}, the total time complexity becomes $O\left(\sum_{m \in \mathcal{M}} \vert \mathcal{V}^{(m)} \vert + \vert \mathcal{E}^{(m)}\vert \right).$

\smallsection{Empirical efficiency.}
{In Appendix~D.3 \cite{online2025appendix}, we empirically compare the inference speed and memory usage of \method against baselines.
The results show that \method does not incur significant computational or memory overhead, compared to the baselines, while achieving the best overall performance.

%% file: 050exp.tex
{\renewcommand{\arraystretch}{1.0}
\begin{table}[t]
\begin{center}
\caption{\label{tab:dataset} Statistics of real-world multi-behavior datasets. 
Note that the Taobao dataset does not have \textit{collect} behaviors.
}
\setlength\tabcolsep{3pt} 
\scalebox{0.93}{
\begin{tabular}{l|cccccc}
    \toprule
    \textbf{Dataset} & \textbf{\#User} & \textbf{\#Item} & \textbf{\#Clicks} & \textbf{\#Collects} & \textbf{\#Carts} & \textbf{\#Buys}\\
    \midrule
    Tmall & 41,738 & 11,953 & 1,813,498 & 221,514 & 1,996 & 225,586 \\
    Taobao & 48,749 & 39,494 & 1,548,162 & -- & 193,747 & 211,022\\
    JData & 93,334 & 24,624 & 1,681,430 & 45,613 & 49,891 & 321,883 \\
    \bottomrule
\end{tabular}}
\end{center}
\end{table}}

\renewcommand{\arraystretch}{0.9}
\begin{table*}[h]
\vspace{-1mm}
\centering
\setlength{\tabcolsep}{2.2pt} 
\caption{RQ1. Performance comparison of both single-behavior and multi-behavior models under the \textbf{standard evaluation settings}. 
\textbf{Boldface} highlights the best performance, and \underline{\smash{underlining}} highlights the second-best performance.
Results are averaged over three runs. Standard deviation values for \method are also reported to indicate its stability. Note that \method consistently outperforms all the methods in all the settings.}
\label{tab:performance_comparison_pro1}
\begin{tabular}{lccccccccccccc}
\toprule
\multirow{2}{*}{\textbf{Dataset}} & \multirow{2}{*}{\textbf{Metric}} & \multicolumn{2}{c}{\textbf{Single-Behavior}} & \multicolumn{9}{c}{\textbf{Multi-Behavior}} \\ 
\cmidrule(lr){3-4} \cmidrule(lr){5-13}
 & & \textbf{MF-BPR} & \textbf{LGCN} & \textbf{LGCN-G} & \textbf{MB-GMN} & \textbf{CML} & \textbf{CIGF} & \textbf{CRGCN} & \textbf{BCIPM} & \textbf{MB-HGCN} & \textbf{MULE} & \textbf{\method(Ours)} \\
\midrule
\multirow{4}{*}{\centering \textbf{Tmall}} 
& \textbf{HR@10}   & 0.0352 & 0.0208 & 0.1358 & 0.0480 & 0.0363 & 0.0443 & 0.1214 & 0.2136 & 0.1991  & \underline{0.2741} & \textbf{0.3764} \std{(0.005)} \\
& \textbf{NDCG@10} & 0.0181 & 0.0102 & 0.0734 & 0.0246 & 0.0181 & 0.0226 & 0.0655 & 0.1109 & 0.1061 & \underline{0.1466} & \textbf{0.1850} \std{(0.007)} \\
\cmidrule(lr){2-14}
& \textbf{HR@20}   & 0.0497 & 0.0307 & 0.1902 & 0.0663 & 0.0548 & 0.0711 & 0.1731 & 0.2983 & 0.2741 & \underline{0.3416} & \textbf{0.5652} \std{(0.005)} \\
& \textbf{NDCG@20} & 0.0216 & 0.0126 & 0.0867 & 0.0282 & 0.0212 & 0.0294 & 0.0782 & 0.1317 & 0.1245 & \underline{0.1629} & \textbf{0.2326} \std{(0.003)} \\
\midrule
\multirow{4}{*}{\centering \textbf{Taobao}}
& \textbf{HR@10}   & 0.0233 & 0.0133 & 0.1432 & 0.0628 & 0.0297 & 0.0644 & 0.0933 & 0.2377 & 0.1392 & \underline{0.2548} & \textbf{0.3371} \std{(0.017)} \\
& \textbf{NDCG@10} & 0.0126 & 0.0071 & 0.0825 & 0.0373 & 0.0149 & 0.0353  & 0.0536 & 0.1276 & 0.0767 & \underline{0.1402} & \textbf{0.1808} \std{(0.013)} \\
\cmidrule(lr){2-14}
& \textbf{HR@20}   & 0.0338 & 0.0197 & 0.1858 & 0.0528 & 0.0507 & 0.0950 & 0.1263 & \underline{0.3276} & 0.1896 & 0.3060 & \textbf{0.4677} \std{(0.018)} \\
& \textbf{NDCG@20} & 0.0152 & 0.0087 & 0.0932 & 0.0210 & 0.0205 & 0.0558 & 0.0608 & 0.1503 & 0.0894 & \underline{0.1639} & \textbf{0.2115} \std{(0.009)} \\
\midrule
\multirow{4}{*}{\centering \textbf{Jdata}}    
& \textbf{HR@10}   & 0.3119 & 0.3095 & 0.4802 & 0.2970 & 0.2302 & 0.3375 & 0.5010 & 0.5532 & 0.5338 & \underline{0.6073} & \textbf{0.6618} \std{(0.005)} \\
& \textbf{NDCG@10} & 0.1885 & 0.1758 & 0.2817 & 0.1734 & 0.1276 & 0.2195 & 0.2905 & 0.3173 & 0.3238 & \underline{0.4300} & \textbf{0.4425} \std{(0.003)} \\
\cmidrule(lr){2-14}
& \textbf{HR@20}   & 0.3826 & 0.3806 & 0.5868 & 0.3859 & 0.3112 & 0.4352 & 0.5850 & 0.6693 & 0.6450 & \underline{0.6790} & \textbf{0.7608} \std{(0.009)} \\
& \textbf{NDCG@20} & 0.2077 & 0.1946 & 0.3078 & 0.1995 & 0.1496 & 0.2423  & 0.3234 & 0.3475  & 0.3533  & \underline{0.4370} & \textbf{0.4687} \std{(0.003)} \\
\bottomrule
\end{tabular}
\end{table*}


\renewcommand{\arraystretch}{0.99}
\begin{table*}[h]
\centering
\setlength{\tabcolsep}{0.9pt} 
\caption{RQ2. Performance comparison of both single-behavior and multi-behavior models under the \textbf{visited- and unvisited-item evaluation settings}. 
\textbf{Boldface} highlights the best performance, and \underline{\smash{underlining}} highlights the second-best performance.
Results are averaged over three runs. Standard deviation values for \method are also reported to indicate its stability.
Note that \method outperforms all the methods in 9 out of 12 evaluation settings.}
\label{tab:performance_comparison_pro2}
\begin{tabular}{lccccccccccccc}
\toprule
\multirow{2}{*}{\textbf{Dataset}} & \multirow{2}{*}{\textbf{Setting}} & \multirow{2}{*}{\textbf{Metric}} & \multicolumn{2}{c}{\textbf{Single-Behavior}} & \multicolumn{9}{c}{\textbf{Multi-Behavior}} \\ 
\cmidrule(lr){4-5} \cmidrule(lr){6-14}
 & & & \textbf{MF-BPR} & \textbf{LGCN} & \textbf{LGCN-G} & \textbf{MB-GMN} & \textbf{CML} & \textbf{CIGF} & \textbf{CRGCN} & \textbf{BCIPM} & \textbf{MB-HGCN} & \textbf{MULE} & \textbf{\method(Ours)} \\
\midrule
\multirow{4}{*}{\centering \textbf{Tmall}} & \multirow{2}{*}{\centering \textbf{Visited}} 
& \textbf{HR@10}   & 0.0288 & 0.3368 & 0.4120 & 0.2740 & 0.2902 & 0.2675 & 0.3906 & 0.4073 & 0.3818  & \underline{0.4545} & \textbf{0.4593} \std{(0.003)}  \\
& & \textbf{NDCG@10} & 0.0151 & 0.1635 & 0.1988 & 0.1234 & 0.1397 & 0.1232 & 0.1852 & 0.1967 & 0.1833 & \underline{0.2211} & \textbf{0.2256} \std{(0.004)} \\
\cmidrule(lr){2-14}
 & \multirow{2}{*}{\centering \textbf{Unvisited}} & \textbf{HR@10}   & \underline{0.0645} & 0.0408 & 0.0526 & 0.0315 & 0.0245 & 0.0269 & 0.0351 & 0.0546 & 0.0432 & 0.0512 & \textbf{0.1024} \std{(0.002)} \\
& & \textbf{NDCG@10} & 0.0215 & 0.0213 & 0.0289 & 0.0167 & 0.0131 & 0.0133 & 0.0181 & \underline{0.0293} & 0.0220 & 0.0273 & \textbf{0.0661} \std{(0.003)} \\
\midrule
\multirow{4}{*}{\centering \textbf{Taobao}} & \multirow{2}{*}{\centering \textbf{Visited}}
& \textbf{HR@10}   & 0.0265 & 0.3782 & 0.5058 & 0.4136 & 0.4029 & 0.3895 & 0.4934 & 0.4935 & 0.4843 & \underline{0.5559} & \textbf{0.5614} \std{(0.019)} \\
& &\textbf{NDCG@10} & 0.0142 & 0.1934 & 0.2557 & 0.1918 & 0.2014 & 0.1814 & 0.2519 & 0.2450 & 0.2464 & \underline{0.2956} & \textbf{0.2985} \std{(0.016)} \\
\cmidrule(lr){2-14}
& \multirow{2}{*}{\centering \textbf{Unvisited}} & \textbf{HR@10}   & 0.0176  & 0.0097 & 0.0140 & 0.0196 & 0.0134 & 0.0139 & 0.0157 & 0.0154 & 0.0197 & \underline{0.0242} & \textbf{0.0272} \std{(0.001)} \\
& & \textbf{NDCG@10} & 0.0097 & 0.0054 & 0.0079 & 0.0101 & 0.0064 & 0.0073 & 0.0082 & 0.0083 & 0.0107 & \underline{0.0137} & \textbf{0.0150} \std{(0.001)} \\
\midrule
\multirow{4}{*}{\centering \textbf{Jdata}} & \multirow{2}{*}{\centering \textbf{Visited}}
& \textbf{HR@10}   & 0.3075 & 0.6935 & 0.7315 & 0.5849 & 0.5678 & 0.5748 & 0.7114 & 0.7278 & 0.7164  & \textbf{0.7605} & \underline{0.7574} \std{(0.001)} \\
& & \textbf{NDCG@10} & 0.1859 & 0.4392 & 0.4499 & 0.2795 & 0.2715 & 0.2781 & 0.4314 & 0.4340 & 0.4325  & \underline{0.4739} & \textbf{0.5088} \std{(0.003)} \\
\cmidrule(lr){2-14}
& \multirow{2}{*}{\centering \textbf{Unvisited}} & \textbf{HR@10}   & 0.3225 & 0.3363 & 0.3655 & 0.2408 & 0.1687 & 0.2276 & 0.2584 & 0.3645 & 0.3910 & \textbf{0.4557} & \underline{0.4159} \std{(0.009)} \\
& & \textbf{NDCG@10} & 0.1948 & 0.1968 & 0.2202 & 0.1500 & 0.0930 & 0.1428 & 0.1393 & 0.2172 & 0.2376  & \textbf{0.3139} & \underline{0.2664} \std{(0.006)} \\ 
\bottomrule
\end{tabular}
\vspace{5pt}
\end{table*}

In this section, we demonstrate the effectiveness of \method.
We address the following four research questions:
\begin{itemize}[leftmargin=*]
    \item \textbf{RQ1 (Overall performance).} \textit{How effective is \method in standard multi-behavior recommendation evaluation settings?}
    \item \textbf{RQ2 (Item type-specific performance).} \textit{How effective is \method in recommending visited and unvisited items, respectively?}
    \item \textbf{RQ3 (Ablation study).} \textit{Are all the key components of \method essential for achieving high performance?}
    \item \textbf{RQ4 (Sensitivity).} \textit{How well does \method perform across various hyperparameter configurations?}
\end{itemize}

\subsection{Experimental Settings}\label{sec:exp:setting}
\smallsection{Datasets.}
We conduct our experiments on the following widely used multi-behavior recommendation datasets:
\begin{itemize}[leftmargin=*]
    \item \textbf{Tmall.\footnote{\url{https://tianchi.aliyun.com/dataset/140281}}} Tmall is an e-commerce dataset from Alibaba.com, containing four user behaviors: click, collect, cart, and purchase. Among these, purchase is considered the target behavior.
    \item \textbf{Taobao.\footnote{\url{https://tianchi.aliyun.com/dataset/649}}} Taobao is an e-commerce dataset from Taobao.com, with three user behaviors: click, cart, and purchase. Among these, purchase is typically considered the target behavior.
    \item \textbf{Jdata.\footnote{\url{https://global.jd.com/}}} JData is an e-commerce dataset from JD.com, containing four user behaviors: click, collect, cart, and purchase. Among these, purchase is considered the target behavior.
\end{itemize}
Some statistics for each dataset are provided in Table ~\ref{tab:dataset}.

\smallsection{Baseline methods.} 
We compare \method against nine baseline methods, including two single-behavior recommendation models and seven multi-behavior recommendation models. 
Specifically, the former consists of MF-BPR~\citep{rendle2012bpr} and LightGCN (LGCN)~\citep{he2020lightgcn}, and the latter consists of LightGCN-G (LGCN-G)~\citep{he2020lightgcn}, MB-GMN~\cite{xia2021graph}, CML~\cite{wei2022constrastive}, CIGF~\cite{guo2023compressed}, CRGCN~\cite{yan2023cascading},  BCIPM~\cite{yan2024behavior}, MB-HGCN~\citep{yan2023mbhgcn}, and MULE~\citep{lee2024mule}.
Among these, MB-HGCN and MULE are the state-of-the-art multi-behavior recommender systems.
We provide further details of these baseline methods in Appendix A.1 \cite{online2025appendix}.


\smallsection{Evaluation metrics.}\label{sec:evalaution_metric}
We measure the evaluation quality with Hit Ratio@K (HR@K) and Normalized Discounted Cumulative Gain@K (NDCG@K), which are metrics commonly leveraged in multi-behavior recommendation studies~\cite{xia2021graph, wei2022constrastive, yan2023cascading, yan2024behavior, yan2023mbhgcn, guo2023compressed, lee2024mule}. For detailed descriptions of these evaluation metrics, please refer to Appendix A.2 \cite{online2025appendix}.


\smallsection{Evaluation protocol details.} In the standard recommendation setting (RQ1), we rank all candidate items for each user based on \(s_{ui}^{*}\) and select the top \(K\) items. To further evaluate the model's performance on specific types of items (RQ2), we also generate separate rankings using $\mathds{1}[(u,i) \in \mathcal{C}_{u}^{(V)}]s^{(V)}_{ui}$ for visited items performance and $\mathds{1}[(u,i) \in \mathcal{C}_{u}^{(U)}]s^{(U)}_{ui}$ for unvisited items performance.
This dual evaluation allows us to assess how effectively the model recommends both types of items. Note that we apply this same evaluation protocol to all baselines for a fair comparison. 


{\renewcommand{\arraystretch}{0.9}
\begin{table*}
\vspace{-1mm}
\caption{\label{tab:ablation_combined} RQ3. Effectiveness of the key components of \method on both visited-item and unvisited-item recommendation. The best performance is highlighted in \textbf{bold}, and the second-best one is \underline{underlined}. H@10: Hit Ratio@10. N@10: NDCG@10.}
\setlength\tabcolsep{5pt} 
\scalebox{1.0}{
\begin{tabular}{l|cc|cc|cc|cc|cc|cc}
    \toprule
    {\textbf{Setting}} & \multicolumn{6}{c|}{\textbf{Visited-item Recommendation}} & \multicolumn{6}{c}{\textbf{Unvisited-item Recommendation}} \\
    \cmidrule(lr){2-7} \cmidrule(lr){8-13}
    \textbf{Datasets} & \multicolumn{2}{c|}{\textbf{Tmall}} & \multicolumn{2}{c|}{\textbf{Taobao}} & \multicolumn{2}{c|}{\textbf{JData}} 
    & \multicolumn{2}{c|}{\textbf{Tmall}} & \multicolumn{2}{c|}{\textbf{Taobao}} & \multicolumn{2}{c}{\textbf{JData}} \\
    \midrule
    \textbf{Metrics} & \textbf{H@10} & \textbf{N@10} & \textbf{H@10} & \textbf{N@10} & \textbf{H@10} & \textbf{N@10} 
    & \textbf{H@10} & \textbf{N@10} & \textbf{H@10} & \textbf{N@10} & \textbf{H@10} & \textbf{N@10} \\
    \midrule
    \textbf{\method-\text{MoE}} & 0.4208 & 0.2063 & 0.5516 & 0.2942 & \underline{0.7549} & 0.5002 
    & 0.0560 & 0.0266 & 0.0218 & 0.0110 & 0.2912 & 0.1832 \\
    \textbf{\method-$\mathcal{L}^{(V)}_{\text{SSL}}$} & 0.4377 & 0.2109 & 0.5185 & 0.2692 & 0.7278 & 0.4735 
    & \underline{0.0896} & 0.0520 & \underline{0.0271} & \underline{0.0148} & 0.4011 & 0.2524 \\  
    \textbf{\method-$\mathcal{L}^{(U)}_{\text{CL}}$} & 0.4423 & 0.2167 & 0.5463 & 0.2898 & 0.7480 & \underline{0.5004} 
    & 0.0765 & 0.0459 & 0.0181 & 0.0101 & \underline{0.4117} & \textbf{0.2665} \\
    \textbf{\method-$\mathcal{L}^{(U)}_{\text{GEN}}$} & \underline{0.4540} & \underline{0.2229} & \underline{0.5538} & \underline{0.2975} & 0.7428 & 0.4975 
    & 0.0854 & \underline{0.0554} & 0.0259 & 0.0140 & 0.3883 & 0.2416 \\ 
    \midrule
    \textbf{\method} & \textbf{0.4593} & \textbf{0.2256} & \textbf{0.5614} & \textbf{0.2985} & \textbf{0.7574} & \textbf{0.5088} 
    & \textbf{0.1024} & \textbf{0.0661} & \textbf{0.0272} & \textbf{0.0150} & \textbf{0.4159} & \underline{0.2664} \\
    \bottomrule
\end{tabular}}
\end{table*}
}

\smallsection{Hyperparameter configurations.}
Note that since we use multiple experts with distinct user and item parameters, an embedding size of \( d = 16 \) in our model corresponds to an embedding size of \( d = 64 \) in the baseline methods.  
Therefore, we tune the embedding size of the baselines within \( d \in \{64, 128\} \) and that of \method within \( d \in \{16, 32\} \) \textbf{to ensure a fair comparison}. 
The batch size and learning rate are fixed to 1024 and $5 \times 10^{-3}$, respectively, across all the methods. 
For the other hyperparameters of the baseline methods, we use the values reported in their respective papers and official GitHub repositories.
For \method, we employ a 2-layer LightGCN~\cite{he2020lightgcn} as the GNN encoder for each behavior graph and the global graph. 
The following hyperparameter search spaces are employed for \method: 
the temperature parameters $\tau$ and $\tau'$ within $\{0.1, 0.2, \cdots, 1.0\}$; the weight for the local-global-aggregation coefficient $\lambda$ within $\{0.4, 0.5, 0.6\}$; 
the loss coefficients $\gamma_{1}$, $\gamma_{2}$ and $\gamma_{3}$ within $\{1e-3, 1e-2, 5e-2, 0.1, 0.5, 1.0\}$.


\subsection{RQ1. Standard Evaluation Settings}\label{sec:rq1}

We compare the performance of \method against those of the baseline methods under the standard evaluation setting, which is widely leveraged in prior studies~\cite{yan2023cascading, meng2023parallel, meng2023hierarchical, yan2024behavior, yan2023mbhgcn, lee2024mule}.
As shown in Table~\ref{tab:performance_comparison_pro1}, \method outperforms all the baseline methods in all settings.
Specifically, compared to the best competitor MULE, \method achieves up to {65.46\%} performance gain in terms of HR@20, 
demonstrating that the general recommendation quality of \method outperforms that of the baseline methods.



\subsection{RQ2. Item Type-Specific Evaluation Settings}\label{sec:rq2}

We evaluate the recommendation quality of \method for visited items and unvisited items, with setting details provided in Section~\ref{sec:exp:setting}.
As shown in Table~\ref{tab:performance_comparison_pro2}, \method outperforms all the baseline methods in 9 out of 12 settings.
Specifically, compared to the second-best method, \method achieves improvements of up to 7.36\% and 125.60\% in terms of NDCG@10 for visited item and unvisited item recommendations, respectively. 
Notably, the gains for the unvisited-item expert are more pronounced, suggesting that our approach is especially effective for the challenging task of recommending unvisited items to users.

\subsection{RQ3. Ablation Study}
We demonstrate the importance of the key components of \method through ablation studies.
To this end, we use four variants of \method, which are:
\begin{itemize}[leftmargin=*]
    \item \textbf{\method-\text{MoE}}: This variant replaces the expert-specific aggregation (i.e., hard gating) with a simple average of the scores from the two experts.
    \item \textbf{\method-\(\mathcal{L}^{(V)}_{\text{SSL}}\)}: This variant excludes the contrastive learning loss (Eq.~\eqref{eq:sslvisited}) for the \textit{visited-item expert}.
    \item \textbf{\method-\(\mathcal{L}^{(U)}_{\text{CL}}\)}: This variant removes the contrastive learning loss (Eq.~\eqref{eq:clunvisited}) for the \textit{unvisited-item expert}.
    \item \textbf{\method-\(\mathcal{L}^{(U)}_{\text{GEN}}\)}: This variant excludes the generative loss (Eq.~\eqref{eq:genunvisited}) used for the \textit{unvisited-item expert}.
\end{itemize}
We evaluate each variant under the same setting of Section~\ref{sec:rq2}. 

As shown in Table~\ref{tab:ablation_combined}, the mixture-of-experts architecture demonstrates effectiveness across all settings. 
Moreover, each specialized self-supervised learning plays a crucial role, as its absence leads to significant performance drops in its expertise setting. 
Specifically, excluding SSL for visited items $(\mathcal{L}^{(V)}_{\text{SSL}})$ results in a notable performance drop in visited item recommendation, while removing SSL for unvisited item $(\mathcal{L}^{(U)}_{\text{CL}}$ and $\mathcal{L}^{(U)}_{\text{GEN}})$ 
leads to a substantial drop in unvisited item recommendation performance. 
These results demonstrate that \method architecture and SSL losses have successfully achieved their design objective.

\subsection{RQ4. Hyperparameter Sensitivity Analysis} 
\begin{figure}[t]
    \centering
        \includegraphics[width=0.925\linewidth]{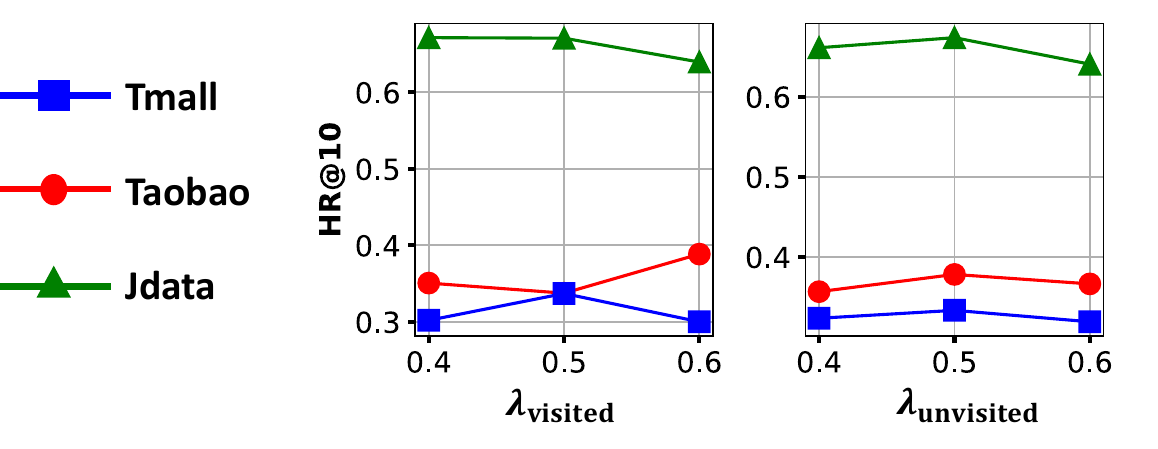}
        \caption{RQ4. Impact of varying each local-global aggregation coefficient $\lambda_\text{visited}$ (for the visited-item expert) and $\lambda_\text{unvisited}$ (for the unvisited-item expert) on Hit Ratio@10.
        }\label{fig:hs_lambda}
\end{figure}

\begin{figure}[t]
         \centering
        \includegraphics[width=0.925\linewidth]{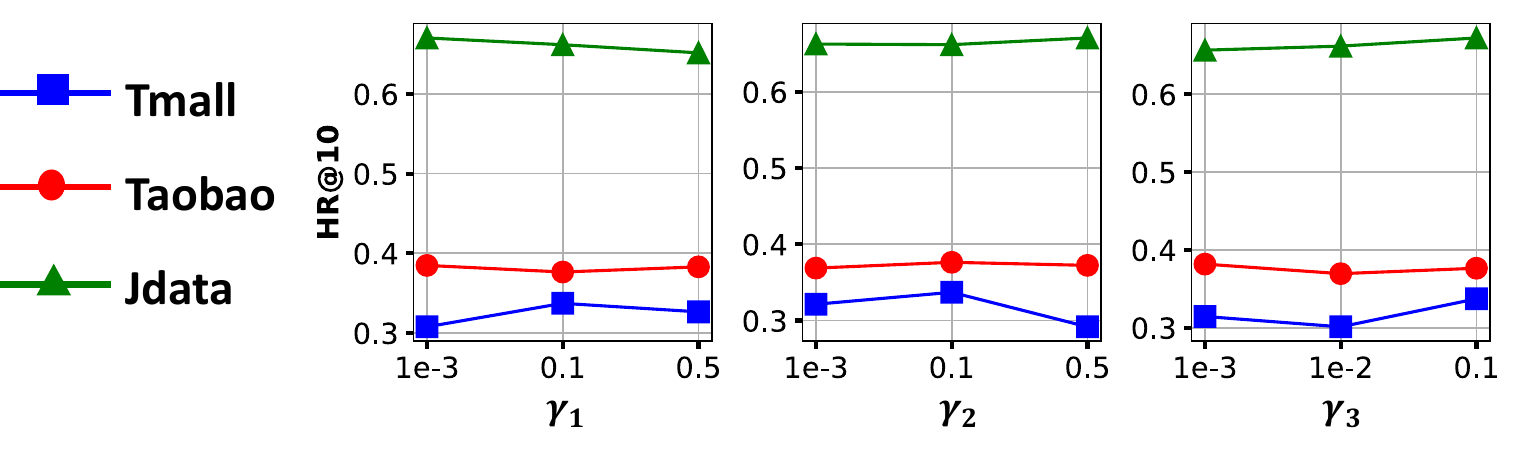}
        \caption{RQ4. Impact of SSL loss weights $\gamma_1$, $\gamma_2$, and $\gamma_3$ on Hit Ratio@10. 
        }\label{fig:hs_loss}
\end{figure}

We evaluate the hyperparameter sensitivity of \method,  examining two key factors: (1) the local–global aggregation coefficient $\lambda_\text{visited}$ and $\lambda_\text{unvisited}$, and (2) the self-supervised learning (SSL) loss weights $\gamma_1, \gamma_2$, and $\gamma_3$. We examine Hit Ratio@10 across various configurations under the standard evaluation setting. As illustrated in Figure~\ref{fig:hs_lambda} and Figure~\ref{fig:hs_loss}, \method consistently demonstrates strong performance under all tested settings. Remarkably, even in its lowest-performing configuration, \method still outperforms the strongest baseline. 

\subsection{Additional Experiments}
In the online appendix~\cite{online2025appendix}, we present additional experiments to further validate the design choices and effectiveness of \method. 
\begin{itemize}[leftmargin=*]
    \item \textbf{Additional metrics.} In Appendix D.1, we present the performance of \method and the baseline methods in terms of additional evaluation metrics (i.e., HR@50 and NDCG@50). 
    Consistent with the results in Section~\ref{sec:rq1}, \method achieves the state-of-the-art recommendation performance.
    \item \textbf{Additional ablation study.} In Appendix D.1, we present additional ablation study results that further demonstrate the effectiveness of our component. 
    \item \textbf{Gating function analysis.} In Appendix D.2, we compare our hard gating approach with soft gating methods, including averaging and or a learnable soft gating function. 
    \item \textbf{Efficiency analysis.} In Appendix~D.3, we empirically compare the inference speed and memory usage of \method against baselines, demonstrating that \method does not incur significant computational or memory overhead, compared to the baselines.
    \item \textbf{Cold-start item evaluation.} In Appendix~D.4, we empirically evaluate the performance on cold-start items, although \method is not explicitly designed for the cold-start scenario.
\end{itemize}



%% file: 070conclusion.tex
In this work, we identify two key challenges in multi-behavior recommender systems: (1) notable performance disparities between visited items (i.e., those a user has interacted through auxiliary behaviors) and unvisited items (i.e., those with which the user has had no such interactions), and (2) difficulty of providing good recommendation for both item types with a single model architecture. 
To tackle these challenges, we propose \method, a mixture-of-experts-based approach where each expert is specialized to recommend either visited or unvisited items. 
This specialization is enhanced by expertise-adaptive self-supervised learning, which aligns each expert’s objective with its designated item type.
Our empirical results demonstrate the superiority of \method over existing methods in both visited-item and unvisited-item recommendations. 
Our code, datasets, and online appendix are available at \url{https://github.com/K-Kyungho/MEMBER}.